# REAL TIME TURBULENT VIDEO PERFECTING BY IMAGE STABILIZATION AND SUPER-RESOLUTION

Barak Fishbain, Leonid P. Yaroslavsky, Ianir A. Ideses

*Dept. of Interdisciplinary Studies, the Iby and Aladar Fleischman Faculty of Engineering, Tel-Aviv University, Tel Aviv 69978, Israel, Phone: 972-3-640-8014, Fax: 972-3-641-0189*

**ABSTRACT**

Image and video quality in Long Range Observation Systems (LOROS) suffer from atmospheric turbulence that causes small neighbourhoods in image frames to chaotically move in different directions and substantially hampers visual analysis of such image and video sequences. The paper presents a real-time algorithm for perfecting turbulence degraded videos by means of stabilization and resolution enhancement. The latter is achieved by exploiting the turbulent motion. The algorithm involves generation of a "reference" frame and estimation, for each incoming video frame, of a local image displacement map with respect to the reference frame; segmentation of the displacement map into two classes: stationary and moving objects and resolution enhancement of stationary objects, while preserving real motion. Experiments with synthetic and real-life sequences have shown that the enhanced videos, generated in real time, exhibit substantially better resolution and complete stabilization for stationary objects while retaining real motion.

## 1. Introduction

Long Range Observation Systems (LOROS) are highly demanded in many fields such as astronomy (i.e. planet exploration), geology (i.e. topographical measurements), ecology, traffic control, remote sensing, and homeland security (surveillance and military intelligence). Ideally, image quality in LOROS would be limited only by the optical setup used, but in reality they suffer from atmospheric turbulence.

The troposphere layer of the atmosphere is in constant motion due to winds and local temperature variations. These variations cause formation of air pockets, which have a uniform index of refraction and can be modeled as spherically shaped turbulent cells in a range of sizes and densities (referred to as "turbulent eddies"). This causes small neighbourhoods in the image sequences to chaotically move in different directions in different frames. As a result, images captured by optical sensors in the presence of atmospheric turbulence are degraded in their resolution and geometry.

Recently, a turbulence compensation algorithm applying local neighborhood methods was introduced suggested [1, 2, 3, 4]. The algorithm uses, for reconstructing distortion-compensated image frames, an adaptive control grid interpolation method based on estimating the local spatial displacement vectors. The algorithm also manages to preserve the genuine motion of the object by evaluating its motion vectors characteristics and making a decision whether to make the correction (turbulent motion) or not (real motion) In the present paper, we describe further improvement of the algorithm which enables full stabilization and resolution enhancement ("super-resolution") of the stationary areas in the field of view under real time implementation, Most super-resolution techniques are based on the fact that no real observation platform can be absolutely stationary. Therefore there are always some micro-movements during the video data acquisition stage. Consequent frames that differ only due to these small movements of the image plane can be combined in order to generate a new image with better spatial resolution. The novel use of the turbulent motion, rather than the acquisition system's vibrations, as the base for super-resolution is presented in this paper.

Generally, the super-resolution process can be divided into 2 main stages. The first is determination, with a sub-pixel resolution, of pixel movements in order to get as many data samples as possible within the sampling interval defined by the sensor geometry. The second is to combine the data observed in every frame in order to generate one new image with better spatial resolution. Super-resolution principles and general multi-channel image recovery are detailed in number of publications. Several researchers treat the problem of high resolution image recovery by designing an efficient multi-frame filtering algorithms, that account for both intra-frame (spatial) and inter-frame (temporal) correlations, for restoring image sequences that are degraded both by blur and noise [5, 6, 7]. Others have formulated solutions to global motion problems, usually from an application perspective [8, 9, 10, 11, 12, 13, 14]. These can be broadly classified as feature-based and flow-based techniques. Feature-based methods extract and match discrete features between frames, and the trajectories of these features are fit to a global motion model. In flow-based algorithms, the optical flow of the image sequence is an intermediate quantity that is used for determining the global motion.

While the turbulent distortions compensation was previously dealt, it was considered to be of an annoyance. The use of turbulence motion for super-resolution was firstly suggested by the authors in [15]. However, the presented method generates a single super-resolved frame from a sequence; hence the motion in the scene is discarded. This is also referred to, in the literature, as static super-resolution (For the definition of static and dynamic super-resolution see [16]).

This paper suggests using turbulent motion for quasi-dynamic super-resolution of turbulent degraded sequences. Quasi-dynamic means that while background areas are stabilized and super-resolved, the moving objects are preserved. The suggested method achieves super-resolution through a flow-based hierarchical mechanism. In the first stage, the scene motion field is extracted. Then, real motion is discriminated from turbulence caused motion in the observed scene. This is described in section 2. Super-Resolution is applied on areas, which contain turbulent motion. This is detailed in section 3.

## 2. Motion Extraction and Discrimination

Extraction of the motion field of a video sequence is carried out in two phases. In the first, a reference image, estimating the stable scene is generated $\bar{I}_{(\hat{x},\hat{y})}$. Then, for each pixel in the original frame $(I_{x,y})$, its coordinates $(\hat{x},\hat{y})$ in the reference frame are determined.

### 2.1 Estimation of the Stable Scene

The reference image is an estimate of the stable scene. Such an image has to be obtained from the input sequence itself [17,18,19]. In order to achieve optimal results, the reference image should have the following properties:

- The reference image should contain only the static background with no moving objects in it.
- It should contain no turbulent induced geometric distortion.
- It should have high signal to noise ratio.

The suggested algorithm for generating the reference image uses a pixel-wise rank filtering in a temporal sliding window. The use of rank smoothing filters such as median and alpha-trimmed mean secures that estimation of the pixel's real value (if there were no turbulence) is close to the mean of the array of the same pixel's values in a long period of time [19], and at the same time that moving objects, whose pixel gray levels form the tails of the gray level distribution in a long sequence, will be eliminated from the estimation. More general element-wise rank filtering in the time sequence to obtain the reference image is used in [1-4]. The suggested method utilizes pixel-wise temporal median filter for extraction of the reference stable frame [20, 21]. **Figure 1**(a) presents a frame extracted from a real-life turbulent degraded video sequence [22], while figure (b) is the reference frame computed by applying element-wise temporal median filtering over 117 frames. The reference frame exhibits an estimate of the stable scene omitting moving objects from the scene.

### 2.2 Motion Field Extraction

The mapping of one turbulent image to a stable image can be obtained by registering a spatial neighborhood, surrounding each pixel in the image, to a reference image. In this way, a field of motion vectors is obtained. Such a registration can be implemented using optical flow methods [23,24,25,26,27,28,29]. In its simplest form, the optical flow method assumes that it is sufficient to find only two parameters of the translation vector for every pixel.

The vectorial difference between the pixel's location in the original image $(I_{(x,y)})$ and its location in the reference image $(\hat{I}_{(\hat{x},\hat{y})})$ is the motion vector $[\Delta\hat{x}, \Delta\hat{y}] = [x-\hat{x}, y-\hat{y}]$.

For the subsequent processing stages, the translation vector is presented in polar coordinates, hence magnitude and angle $\{M_{(x,y,t)}, \theta_{(x,y,t)}\}$ of the motion vector. Having the motion field in hand, one can discriminate real motion from turbulent one through a statistical analysis of the Magnitude $\{M_{(x,y,t)}\}$ and Angle $\{\theta_{(x,y,t)}\}$ components of the motion field.

### 2.3 Motion Discrimination

In order to avoid, in course of the generation of the super-resolved frame, integration of irrelevant data, pixels that represent real moving objects must be extracted from the observed frames. Hence a Real Motion Separation Mask ($RMSM_{(x,y,t)}$) has to be formed for each pixel in each incoming frame. To this end, a real-time two-stage decision mechanism is suggested [4]. The first step is aimed at extracting areas, such as background, that are most definitely stationary. It is designed to achieve fast real time computation. In most cases a great portion of the stable parts in the scene will be extracted at this stage. The rest of the pixels are dealt with at the second step. At this phase, the gray-level difference between running value of each pixel of the incoming frame and its temporal median is calculated as "real-motion measure". **Figure 1**(c) represents in darker pixels, pixels, which were tagged as real-motion. While this first stage detects most of the background pixels as such, it presents estimation noise. This noise is eliminated at the second stage.

The second step improves extraction accuracy at the expense of higher computational complexity, but it handles a substantially smaller portion of the pixels. This stage uses computing and statistical analysis of optical flow for motion segmentation ([1-4]). Cluster analysis of the Magnitude distribution function for all (*x, y*), in a particular frame, allows separating two types of motion amplitudes: small and irregular and large and regular. The first is associated with small movements caused by turbulence. The latter corresponds to movements caused by real motion. Pixel's motion discrimination through angle distribution is achieved by means of statistical filtering of the angle component motion field. For each pixel, its neighborhood's angles' standard deviation is computed. Turbulent motion has chaotic directions. Therefore, a motion filed vectors in a small spatial neighborhood distorted by turbulence has considerably large angular standard deviation. Real motion, on the other hand, has strong regularity in its direction and therefore its angles' standard deviation value over a local neighborhood will be relatively small. Homogeneous background areas contain no motion. Therefore the standard deviation of the zero motion vectors will be zero as well. **Figure 1**(d) represents the pixels which are tagged as containing real motion in darker pixels. While the moving car is preserved, the estimation noise presented by the first estimation phase is eliminated.

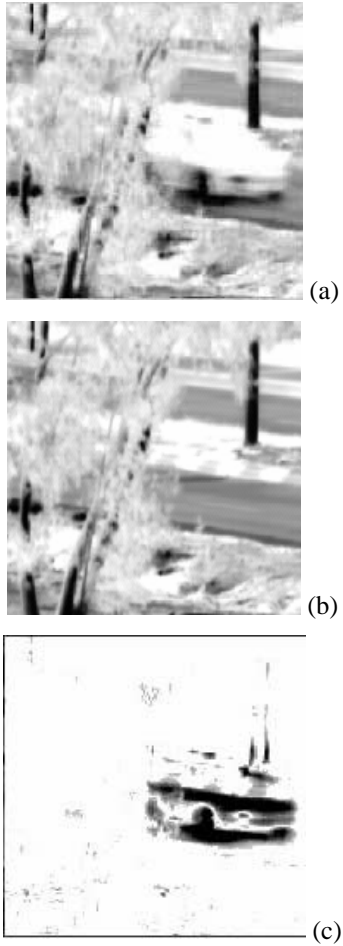

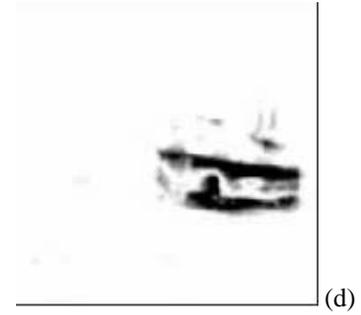

Figure 1 – Motion Extraction and Discrimination. (a) is a frame extracted from a real-life turbulence degraded sequence. (b) is the reference frame computed over 117 frames by applying element-wise temporal median. Darker pixels in (c) represent pixels in which real motion was detected through computation of the absolute gray-level difference of the processed frame from the reference frame and is aimed at extracting areas, such as background, that are most definitely stationary. Darker pixels in (d) represent pixel tagged as containing real motion by statistically analyzing the motion field. The moving car is preserved; the estimation noise presented by the first estimation phase is eliminated.

### 3. Generation of Super-Resolved Frames

The motion vector maps for all frames are used to enhance the resolution of the acquired fames. This is achieved by replacement of pixels in an interpolated estimation of the steady scene by pixels from the frames according to their position defined by their motion vector with sub-pixel accuracy.

#### 3.1 Accumulation of Background or Stationary Pixels' Information

Let the processed sequences frames' size be $(K \times L)$ pixels. The super-resolved sequence frames' size is $(N \cdot K \times N \cdot L)$, where $(N)$ is governed by the user. As described in section 2.3, for every pixel in an incoming frame, the translation vector $(\Delta \hat{x}_{x,y}, \Delta \hat{y}_{x,y})$ is used to decide whether the pixel is of a background or real-moving object. Given that a pixel $(x, y)$ is tagged as background or stationary, the corresponding gray-level value in the super-resolved frame is given by:

$$I(x^{SR}, y^{SR}) = I(x, y) \qquad \text{Eq. 1}$$

where $x^{SR}$ and $y^{SR}$ indicate the pixel's location in the supper-resolved frame and are given by:

$$\{x^{SR}, y^{SR}\} = \{round[N \cdot (x + \Delta \hat{x})], round[N \cdot (y + \Delta \hat{y})]\} \qquad \text{Eq. 2}$$

where $(x, y)$ is the pixel's location in the acquired frame and $(\Delta \hat{x}_{x,y}, \Delta \hat{y}_{x,y})$ is its translation vector computed with a sub-pixel accuracy.

Throughout the supper-resolution data accumulation process there are three possible scenarios: (i) a certain pixel is assigned with a value once; (ii) a certain pixel is assigned with a value more than once; (iii) a certain pixel is not assigned with any value. The first scenario presents no problem; the single intensity gray-level value will be the output in the super-resolved sequence. For coping with the second scenario, one has to integrate the multiple values into one output gray-level value. This can be achieved through applying an average or a median filters on all accumulated data samples. In the case where no data is available at all for a certain pixel, the interpolated stable scene estimate is used as approximation of the super-resolved image. For image interpolation discrete sinc-interpolation, as the numerically optimal interpolation method, is used [30, 31, 32, 33]. **Figure 2**(a) shows the super-resolved frame generated from a real-life turbulent degraded sequence (see **Figure 1**(a) and [22]). While the resolution is enhanced, the motion is omitted from the output sequence. The resolution enhancement is noticeable in **Figure 2**(b) and (c), where the left-hand side is a fragment taken from the reference frame, while the right-hand side is the super-resolved corresponding segment. The extracted segments are marked on (a).

### 3.2 Generation of the Output Frames with Real-Motion Preservation

Based on the notations derived in the previous sections, .the output frame $F_{(x,y,t)}$ is given by:

$$F_{(x,y,t)} = \bar{I}^{SR}_{(x,y,t)} \bullet \{1 - RMSM_{(x,y,t)}\} + I_{(x,y,t)} \bullet RMSM_{(x,y,t)} \quad \text{Eq. 3}$$

where $\bar{I}^{SR}_{(x,y,t)}$ is the super-resolved frame of the background and stationary objects; $I_{(x,y,t)}$ is the current processed frame ($t$); $RMSM_{(x,y,t)}$ is real motion separation mask which is described in 2.3 and "$\bullet$" denotes element-wise matrix multiplication. **Figure 2**(c) illustrates the qusi-dynamic super-resolution output. It presents the same super-resolved frame as presented in figure (a), while real-motion is present in the scene.

### 4. Conclusion

The paper presents an algorithm for real time stabilization and resolution enhancement through atmospheric turbulence distortions in video sequences while keeping the real moving objects in the video unharmed. The algorithm is based on three building blocks: (1) estimation of the stable scene, (2) real motion extraction, and (3) background and stationary objects resolution enhancement and generation of the output frames. Moving objects are located and the resolution enhancement is applied only to the static areas of images. To this goal, for each pixel in the incoming frame it is decided whether it is of a moving or a stationary object. A hierarchical two-stage decision making mechanism is suggested to this goal. At the first stage, the absolute difference of the pixel's gray-level value and the temporal median presenting is used to generate the motion extraction mask. This stage is computationally light and it allows to extracts most of stationary areas. The second stage improves accuracy of separating moving objects by more computationally complex algorithms. At this stage, optical flow computation is used for motion segmentation. Discriminating real motion is achieved through statistical analysis of the magnitude and angle of the motion field elements, which result in t Real Motion Separation Mask, RMSM. Finally, all areas in the incoming frame, which were tagged as stationary, are integrated into a super-resolution process

Experiments with real-live video sequences show that the restored videos exhibit excellent stability for stationary objects and yet retain the moving objects unharmed and easier to visually detect and track in a stable higher resolution background. The super-resolved image is a better candidate then the original one for further image processing tools such as aperture correction and noise filtering.

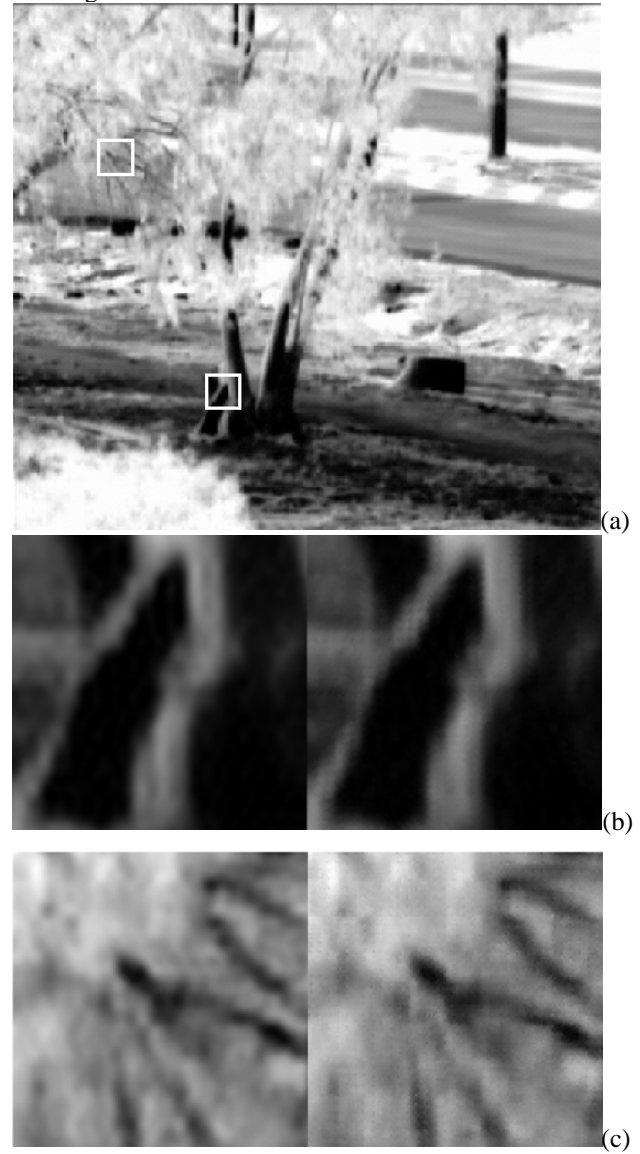

(a)

(b)

(c)

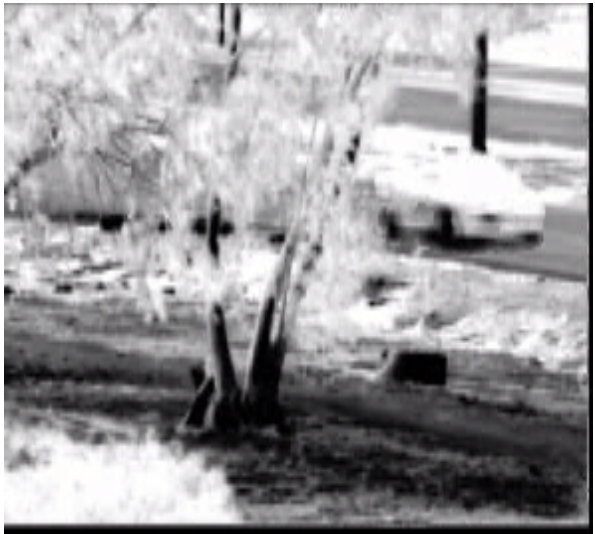

(c)

Figure 2. Static and Quasi-Dynamic Super-Resolution. Figure (a) is a super-resolved frame generated from a real life sequence (see Figure 1(a) and [22]). Images (b) and (c) illustrate the resolution enhancement. On both, the left-hand side is a fragment taken from the reference frame, while the right-hand side is the corresponding super- resolved segment. The corresponding fragments are marked on (a). (d) is the same super-resolved frame as (a), while real-motion is preserved.